\documentclass[conference]{IEEEtran}
\usepackage{cite}
\usepackage{amsmath,amssymb,amsfonts}
\usepackage{algorithmic}
\usepackage{graphicx}
\usepackage{textcomp}
\usepackage{xcolor}
\usepackage{amsmath}
\usepackage{amssymb}
\usepackage{cite}
\usepackage{amsmath,amssymb,amsfonts}
\usepackage{algorithmic}
\usepackage{graphicx}
\usepackage{textcomp}
\usepackage{xcolor}
\usepackage{amsmath}
\usepackage{amssymb}

\makeatletter
 \let\old@ps@headings\ps@headings
 \let\old@ps@IEEEtitlepagestyle\ps@IEEEtitlepagestyle
 \def\confheader#1{%
 \def\ps@headings{%
 \old@ps@headings%
 \def\@oddhead{\strut\hfill#1\hfill\strut}%
 \def\@evenhead{\strut\hfill#1\hfill\strut}%
 }%
 \def\ps@IEEEtitlepagestyle{%
 \old@ps@IEEEtitlepagestyle%
 \def\@oddhead{\strut\hfill#1\hfill\strut}%
 \def\@evenhead{\strut\hfill#1\hfill\strut}%
 }%
 \ps@headings%
 }
 \makeatother

\makeatletter
 \let\old@ps@headings\ps@headings
 \let\old@ps@IEEEtitlepagestyle\ps@IEEEtitlepagestyle
 \def\confheader#1{%
 \def\ps@headings{%
 \old@ps@headings%
 \def\@oddhead{\strut\hfill#1\hfill\strut}%
 \def\@evenhead{\strut\hfill#1\hfill\strut}%
 }%
 \def\ps@IEEEtitlepagestyle{%
 \old@ps@IEEEtitlepagestyle%
 \def\@oddhead{\strut\hfill#1\hfill\strut}%
 \def\@evenhead{\strut\hfill#1\hfill\strut}%
 }%
 \ps@headings%
 }
 \makeatother

 \usepackage[pscoord]{eso-pic}

\newcommand{\placetextbox}[3]{
 \setbox0=\hbox{#3}
 \AddToShipoutPictureFG*{ \put(\LenToUnit{#1\paperwidth},\LenToUnit{#2\paperheight}){\vtop{{\null}\makebox[0pt][c]{#3}}}
 }
 }

 \placetextbox{.23}{0.055}{\small{978-1-6654-5305-9/22/\$31.00~\copyright 2023 IEEE}}

\def\BibTeX{{\rm B\kern-.05em{\sc i\kern-.025em b}\kern-.08em
    T\kern-.1667em\lower.7ex\hbox{E}\kern-.125emX}}
 
\begin{document}

\title{Leveraging a Randomized Key Matrix to Enhance the Security of Symmetric Substitution Ciphers}

\author{\IEEEauthorblockN{Shubham Gandhi\textsuperscript{1},
    Om Khare\textsuperscript{1},
    Mihika Dravid\textsuperscript{1},
    Mihika Sanghvi\textsuperscript{1},
    Sunil Mane\textsuperscript{1},\\
    Aadesh Gajaralwar\textsuperscript{2},
    Saloni Gandhi\textsuperscript{3}
}

\IEEEauthorblockN{\textsuperscript{1} Department of Computer Science and Engineering\\
COEP Technological University\\
Pune, Maharashtra, India}
\IEEEauthorblockN{\textsuperscript{2} Department of Computer Engineering\\
BRACT's Vishwakarma Institute of Information Technology\\
Pune, Maharashtra, India}
\IEEEauthorblockN{\textsuperscript{3}Independent Researcher}

}

\maketitle

\begin{abstract}
%The security of information in digital communication is of paramount importance, and encryption techniques play a vital role in ensuring the confidentiality of sensitive data. This paper explores a novel approach to enhancing the security of substitution ciphers, a classical encryption technique, by introducing a randomized key matrix for various file types, including binary and text files. Although substitution ciphers hold historical significance, they remain susceptible to various cryptanalysis techniques, such as frequency analysis and known-plaintext attacks.The polyalphabetic substitution approach proposed in this study introduces a novel enhancement method to create a randomized key matrix with dimensions of 95x94. The key employed for encryption and decryption is limited to a maximum of 16 characters and can encompass any ASCII letter. This code table serves as the foundation for each encryption and generates a unique random key consisting of ASCII characters and their corresponding ciphertext. Extensive experimentation and evaluation, such as frequency analysis and known-plaintext attacks, as well as speed analysis, showcase the effectiveness of the proposed method in enhancing the security of traditional substitution methods for file encryption and decryption.

An innovative strategy to enhance the security of symmetric substitution ciphers is presented, through the implementation of a randomized key matrix suitable for various file formats, including but not limited to binary and text files. Despite their historical relevance, symmetric substitution ciphers have been limited by vulnerabilities to cryptanalytic methods like frequency analysis and known plaintext attacks. The aim of our research is to mitigate these vulnerabilities by employing a polyalphabetic substitution strategy that incorporates a distinct randomized key matrix. This matrix plays a pivotal role in generating a unique random key, comprising characters, encompassing both uppercase and lowercase letters, numeric, and special characters, to derive the corresponding ciphertext. The effectiveness of the proposed methodology in enhancing the security of conventional substitution methods for file encryption and decryption is supported by comprehensive testing and analysis, which encompass computational speed, frequency analysis, keyspace examination, Kasiski test, entropy analysis, and the utilization of a large language model.

\end{abstract}

\begin{IEEEkeywords}
Encryption, Substitution Cipher, Randomized Key Matrix, Keyspace, Frequency Analysis, Kasiski Analysis, Decryption.
\end{IEEEkeywords}
\section{Introduction}
%In the fast-paced expansion of digital communication and electronic data exchange today, many of us engage in online communication without considering its security implications. We often share our confidential information and personal secrets in cyberspace. Whether we approve of it or not, our online activities leave a digital footprint in cyberspace. To be exact, the information we transmit is frequently unguarded and vulnerable to manipulation by cybercriminals. This emphasizes the critical requirement for modern cryptography to identify means of safeguarding sensitive or confidential information. The security of cyberspace hinges on the efficient encryption and decryption of data. As such, we need to convert information into an unreadable format so that it can be protected and accessed only by those authorized to do so. Cryptography plays a pivotal role in safeguarding information communicated through computers. It involves transforming data into an indecipherable format, ensuring that only the intended recipient can comprehend and utilize it. Essentially, cryptography is the science and art of concealing essential and confidential information to prevent unauthorized individuals from accessing it.

\subsection{Problem Statement}
Cryptography plays an instrumental role in securing information communicated through computer systems, involving the conversion of data into an unintelligible format, thereby ensuring that only the intended recipients possess the ability to decipher and utilize it. Despite their historical significance, classical substitution ciphers such as Caesar, Playfair, and Vigenère can be targeted by various cryptanalysis and are vulnerable to modern-day cyber threats. The need arises to explore innovative strategies for enhancing the security of substitution ciphers while maintaining their elegance and accessibility. The objective of this paper is to address the limitations of traditional substitution ciphers by proposing the integration of a randomized key matrix—a novel approach that enhances security and updates the traditional method to align with contemporary demands.
\subsection{Motivation}
Complex algorithms such as Advanced Encryption Standard (AES), Data Encryption Standard (DES), Triple Data Encryption Standard (3DES), Rivest Cipher 6 (RC6), Rivest Cipher 2 (RC2), and Blowfish have been yielded by modern cryptographic advancements. Motivated by the critical need to address the vulnerabilities inherent in classical substitution ciphers such as frequency analysis attacks, small keyspace, and known-plaintext attacks, while harnessing their simplicity and historical value, this research aims to enhance the security of substitution ciphers while retaining their simplicity and efficiency.

\subsection{Objectives}
The development of a novel approach to enhance the security of substitution ciphers, by integrating a randomized key matrix to address the vulnerabilities inherent in conventional substitution ciphers is the primary objective of this research. Their resistance against various sophisticated attack strategies is intended to be bolstered. Additionally, an exploration of existing cryptography techniques and modern enhancements to substitution cipher techniques is undertaken through a comprehensive literature review.

\begin{figure*}[ht]
  \centering
  \includegraphics[width=\textwidth]{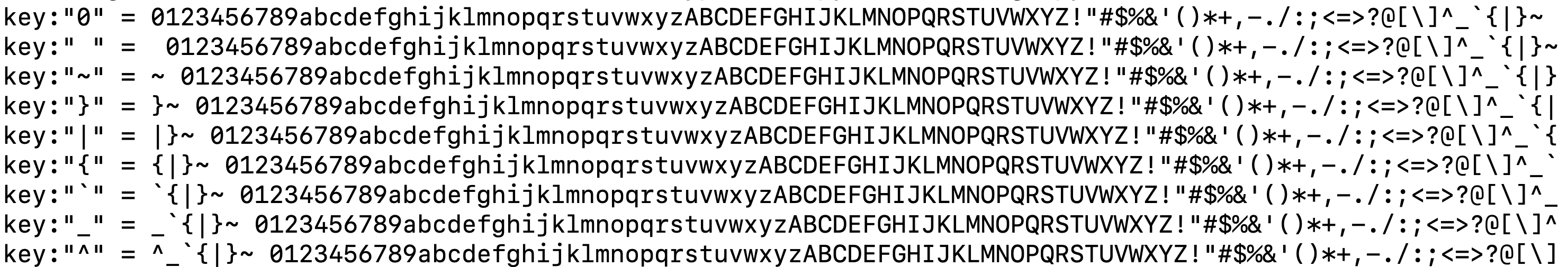}
  \caption{Subset of the Key Matrix}
  \label{fig:key_matrix}
\end{figure*}

\section{Related Work}
Cryptography plays a vital role in safeguarding computer communications by transforming data into an unintelligible format, preventing unauthorized access, and ensuring efficient and secure data transfers.
Within cryptography, Symmetric and Asymmetric\cite{0} methodologies are the two recognized categories.\cite{1}\cite{sym} A critical factor in ensuring communication security is the key size, which varies between the two approaches. Symmetric cryptography employs smaller keys than asymmetric cryptography, making it potentially less secure for highly sensitive data.\cite{4} The encryption process utilizes the American Standard Code for Information Interchange (ASCII) values of the data to be encrypted while introducing a modified string as a shared key for both encryption and decryption. This single key is used for both procedures, with minor adjustments, and is applicable when the input and key lengths are the same. While asymmetric cryptography requires more computational time than symmetric cryptography, making it complex for large data volumes, its larger key size leads to a two-step process where public key cryptography is first used for key exchange, followed by encryption or decryption via symmetric key cryptography \cite{2}\cite{3}. The computational time in cryptography includes encryption, decryption, key generation, and key exchange, with encryption and decryption time linked to the conversion of plaintext to ciphertext and ciphertext to plaintext respectively. The duration of key generation varies based on key length. Key exchange time relies on the communication channel between sender and receiver.

Substitution ciphers represent a fundamental category of encryption methods that operate by substituting one element for another according to predefined rules. Extensive research and enhancements have been conducted on the classical Caesar cipher.\cite{6} A hybrid version of the classical and modern cipher properties has been proposed to improve the Caesar cipher, incorporating a random number generation technique for key generation operations. This extended Caesar cipher includes alphanumeric characters and the key used for the Caesar Substitution is derived using a key Matrix Trace value restricted to Modulo 94. The Matrix elements are generated using a recursive random number generation equation. The second stage of encryption has been performed using columnar transposition with arbitrary random order column selection, significantly enhancing resistance to brute force attacks.
Another proposal\cite{7} presents a modified hybrid of the Caesar cipher and Vigenère cipher to increase the diffusion and confusion characteristics of the ciphertext by incorporating techniques from modern ciphers such as XORing the key to the first letter of the plain text and proceeding iteratively with each subsequent character. \cite{8} An enhancement to Ceasar cipher in which the method of encryption depends on the position of the bit in the message, where the sender transposes the bits by shifting characters in odd positions to the left and characters in even positions to the right.
\\
The Vigenère cipher, \cite{9} a substitution cipher, works by selecting a keyword that determines the shifts for the encryption process. Each letter of the keyword corresponds to a shift value. The plaintext letters are shifted by these values cyclically to create the ciphertext \cite{10}. Additionally, a cryptosystem based on the Vigenère cipher has been proposed, utilizing a multilevel encryption scheme. It involves using a Vigenère table to obtain a new ciphertext from an equivalent fixed-length plaintext and key. This new ciphertext then becomes a key for encrypting the original plaintext, creating a final ciphertext. The decryption process is reversed by the receiver \cite{12}.
\\
Shifting focus to the Playfair cipher, \cite{13} this method operates on character blocks by converting plaintext diagrams to ciphertext diagrams using a pre-shared key. Further modifications of the Playfair cipher such as utilizing different matrix sizes for the key. Comparative analysis indicates that while the size of the plaintext has minimal effect, larger key sizes lead to stronger encryption, and increasing the matrix size enhances encryption results. Incorporating hybrid approaches, \cite{14} combines the Vigenère and columnar transposition ciphers. The hybrid technique applies a columnar transposition cipher to the plaintext using a randomly chosen key to generate an initial ciphertext. This ciphertext then becomes the key for the subsequent Vigenere encryption. This approach adds an extra layer of complexity, enhancing the security of the final ciphertext against existing cryptanalysis methods.

\begin{figure*}[ht]
  \centering
  \includegraphics[width=\textwidth]{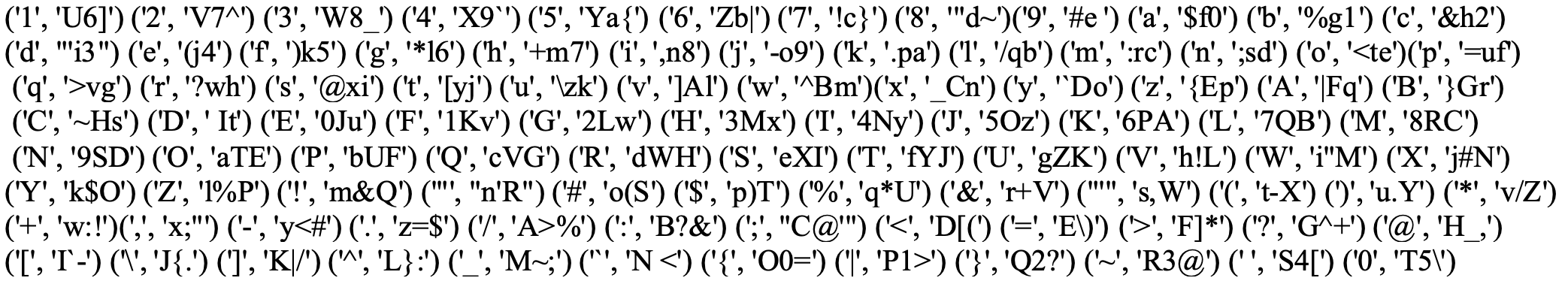}
  \caption{Reference for substitution}
  \label{fig: KeySet}
\end{figure*}

\section{Methodology}
A comprehensive methodology has been developed for our proposed symmetric encryption technique, addressing the limitations of classical substitution ciphers such as their limited keyspace and vulnerability to cryptanalysis. The methodology encompasses a series of steps, namely, key matrix generation, random key generation, encryption, and decryption. In the case of encryption, input is provided in the form of plaintext and key length, while during decryption, input consists of ciphertext and key. The length of the key is subject to variation from 1 to a maximum of 16, implying an inversely proportional relationship between security and computational speed.
\subsection{Key Matrix Generation}
A key matrix with dimensions of 95x94 is created by the algorithm, utilizing a diverse range of ASCII characters such as upper and lowercase alphabets, and numeric and special characters. This matrix serves as the foundation for our proposed methodology. While the first row of the key matrix remains constant, subsequent rows are sequentially populated following a specific pattern. In each row, the initial character functions as a unique identifier for that particular row in the key matrix. A subset of the key matrix is illustrated in Figure \ref{fig:key_matrix}.

\subsection{Key Generation Process}
The flexibility to select an encryption key length is offered by the algorithm, with a maximum limit of 16 characters. This constraint strikes a balance between security and computational efficiency. The process of generating a random encryption key involves the following sequential steps:
\begin{enumerate}

\item \textit{Row Selection from Key Matrix:} A specific number of rows from the key matrix are randomly picked by the algorithm. The number of chosen rows corresponds to the selected key length. The foundation of the encryption key is formed by these selected rows.

\item \textit{Substitution with Unique Sets:} Within the selected key rows, a set is formed by combining one character from each of the chosen rows, and this set is utilized to substitute a single character in the plaintext. The size of this set corresponds to the selected key length. The algorithm's overall security is significantly enhanced by this polyalphabetic substitution mechanism.
\end{enumerate}
By implementing character substitutions in this manner, several security benefits are yielded by the algorithm. An added layer of complexity is introduced by the non-linear and non-deterministic approach to substitutions, which considerably increases the difficulty of the decryption process for potential attackers. Furthermore, the algorithm's resilience against brute-force attacks is substantially fortified through the utilization of diverse characters in conjunction with a sizable key matrix.

\subsection{Encryption}

Using the newly established character sets as a reference for encryption as illustrated in Figure \ref{fig: KeySet}, the index of each plaintext character within this custom set is identified by the algorithm. Subsequently, the identified plaintext character is replaced by its corresponding set of encrypted characters. After each character in the plaintext is replaced with a set of characters where the size of the set is the key length, the corresponding cipher text is formed, and the key is the first character of the randomly selected rows in the key matrix.

\begin{figure}[ht]
  \centering
  \includegraphics[width=\columnwidth]{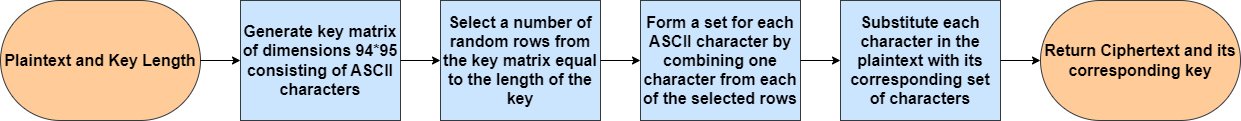}
  \caption{Encryption Workflow}
  \label{fig:ew}
\end{figure}

In essence, from the set of selected rows, each individual letter in the plaintext undergoes substitution. This substitution process involves creating a combination of characters derived from all the selected rows. The resulting encrypted character, stemming from this multi-row combination, is then integrated into the ciphertext.The encryption workflow is illustrated in Figure \ref{fig:ew}.

\subsection{Decryption}
The decryption process involves the reversal of the encryption process. It is initiated by utilizing the encryption key to identify the rows within the key matrix. Each row is distinguishable by its initial character, which serves as a unique identifier. The content of these identified rows is combined to retrieve the original character corresponding to the encrypted set of characters in the cipher text, facilitated by the algorithm.

\begin{figure}[ht]
  \centering
  \includegraphics[width=\columnwidth]{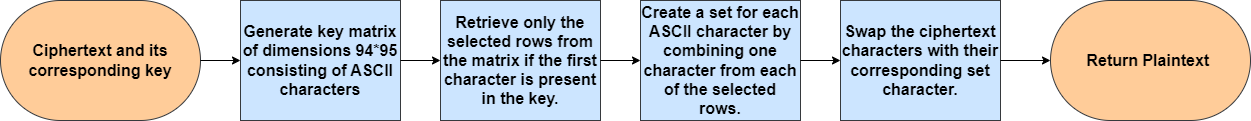}
  \caption{Decryption Workflow}
  \label{fig:dw}
\end{figure}

Upon the recovery of this original character, it is incorporated into the expanding plaintext. This process is iterated for every encrypted character in the ciphertext, gradually reconstructing the original plaintext. The decryption workflow is illustrated in Figure \ref{fig:dw}.

\section{Experimental Evaluation}
The experimental evaluation section of this research constitutes a comprehensive exploration of the proposed encryption technique's capabilities and resilience. Through a series of meticulously designed experiments, we delve into various aspects of the encryption method's performance and security. From computational speed comparisons with traditional substitution ciphers to rigorous frequency analysis, keyspace examination, and robustness evaluation against known cryptanalysis techniques, our experimental approach aims to provide a holistic understanding of the proposed encryption technique's strengths and limitations. Furthermore, we extend our investigation to include challenges posed by advanced technologies, such as Large Language Models (LLMs), as we analyze their impact on the security of our encryption method. 
\subsection{Dataset Preparation}
In the context of our experimental evaluation, a diverse dataset was prepared, encompassing various types of texts such as plain English sentences, technical documents, and excerpts from literary works. The objective of the dataset is to capture real-world scenarios in which the security of text-based information assumes importance.

\subsection{Implementation Details}
The proposed substitution cipher algorithm was implemented as outlined in the methodology. The key length was designated as 8 characters, aiming to strike a balance between security and computational efficiency. As a basis for comparison, classical substitution cipher algorithms including Ceasar, Playfair, and Vigenère were implemented, wherein each character within the plaintext was replaced with a predetermined character from a singular key.
\begin{figure}[ht]
  \centering
  \includegraphics[width=\columnwidth]{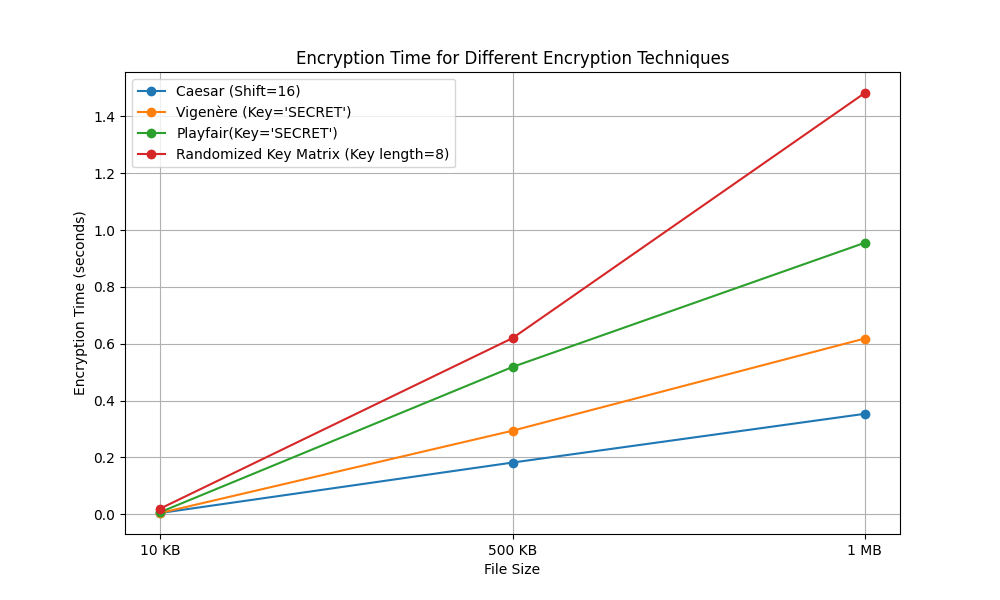}
  \caption{Comparison of Encryption Speeds for Various Substitution Ciphers}
  \label{fig:et}
\end{figure}
\begin{figure}[ht]
  \centering
  \includegraphics[width=\columnwidth]{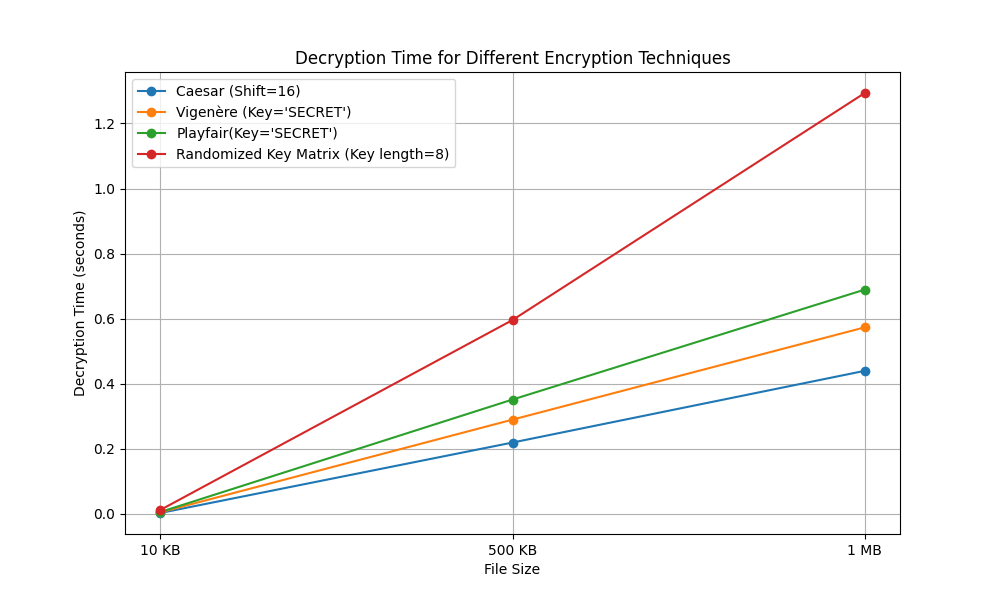}
  \caption{Comparison of Decryption Speeds for Various Substitution Ciphers}
  \label{fig:dt}
\end{figure}

\subsection{Computational Speed Analysis}
 Utilization of a laptop equipped with a 2.80GHz CPU was executed to assess encryption time, and decryption time  \cite{speed}\cite{15.5} across a spectrum of file sizes—namely, 10KB, 500KB, and 1MB. The prescribed key size of 8 was maintained. Competitive speeds are exhibited by our approach, even in comparison to simpler substitution techniques such as Caesar, Vigenère, and Playfair, particularly for diminutive file sizes. The necessary time was computed using (\ref{eq:accuracy}). The Encryption and Decryption time for various substitution ciphers are illustrated in Figure \ref{fig:et} and Figure \ref{fig:dt} respectively.
\begin{equation}
\textit{Computation Time} = {\textit{End Time}}-{\textit{Start Time}} 
\label{eq:accuracy}
\end{equation}
where the computation time is computed by subtracting the start time from the end time

\subsection{Frequency Analysis}
\begin{figure}[ht]
  \centering
  \includegraphics[width=\columnwidth]{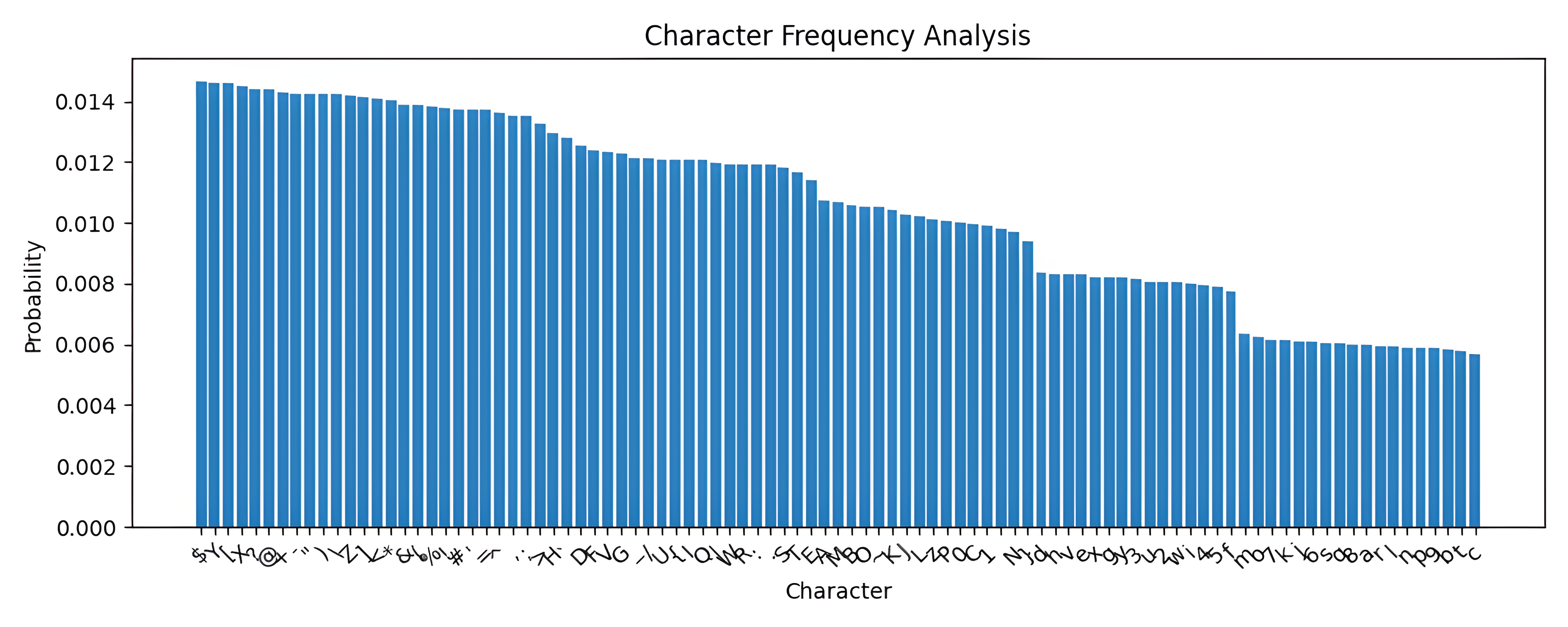}
  \caption{Character Frequency Analysis}
  \label{fig:fa}
\end{figure}
Insightful results have been yielded by the conducted frequency analysis  \cite{freq} on the ciphertext generated through the application of our proposed substitution cipher method. Upon analyzing the histogram of character occurrences, the uniform distribution indicates a polyalphabetic nature and a diverse range of characters, which spans uppercase and lowercase letters, special characters, and numeric characters. The non-uniform distribution of character probabilities further emphasizes this intricacy, revealing a nuanced transformation process. Remarkably, characters such as `v' and 'B' emerge with the highest probabilities, while comparable probabilities among characters like `y', `@', `z', and `?' highlight a balanced usage. It is noteworthy that common English letters appear with lowered probabilities, effectively disrupting conventional language frequencies. Special characters and numeric symbols also play a significant role, thereby confounding frequency analysis expectations. By encompassing a broad range of characters and introducing non-standard frequencies, the capacity of our proposed method to enhance the security of substitution ciphers is showcased, effectively deterring traditional frequency-based attacks. The character frequency distribution graph is illustrated in Figure \ref{fig:fa}.

\subsection {Key Space Analysis}
Keyspace analysis plays a crucial role in the evaluation of cipher security by quantifying the number of potential keys that an attacker must consider for breaking the encryption through brute force. Keyspace is the total number of possible values of keys in a crypto algorithm. A comprehensive analysis of the keyspace strength across four distinct cryptographic algorithms—namely, the classic Caesar cipher, the more advanced Playfair cipher, and our novel encryption scheme—is undertaken. Their corresponding keyspace is outlined in Table \ref{table:keyspace}.

Commencing with the Caesar cipher, the keyspace is found to be limited to a mere 25 possibilities, thus emphasizing its vulnerability to brute-force attacks. In contrast, enhanced security is offered by the Vigenère cipher due to a larger keyspace, which is determined by the length of the secret keyword. The Playfair cipher, leverages a 5x5 letter grid, resulting in an impressive keyspace of 1.55 x 10\textsuperscript{25} unique configurations, rendering it notably resilient against cryptographic attacks.

In comparison, our cipher showcases an extraordinary keyspace of 4.40 x 10\textsuperscript{25}. This keyspace's unprecedented magnitude substantially enhances our cipher's security profile, effectively rendering exhaustive search attacks like brute force practically infeasible.
\begin{table}[htbp]
  \centering
  \caption{Keyspace Analysis}
  \label{table:keyspace}
  \begin{tabular}{|c|c|}
    \hline
    Cipher & Keyspace \\
    \hline
Caesar  & 25 \\
Playfair  & $1.55 \times 10^{25}$ \\
Randomized Key Matrix & $4.40 \times 10^{25}$ \\
    \hline
  \end{tabular}
\end{table}

\subsection{Kasiski Analysis}
A Kasiski test \cite{kas}\cite{kas2} was conducted to simulate an attack and assess the resilience of our approach. The Kasiski test is a potent cryptanalysis technique employed to ascertain the length of the encryption key in polyalphabetic substitution ciphers.

The ciphertext provided below in Figure \ref{fig:ka}, generated by our algorithm with a key length of 6, is taken into consideration.
\begin{figure}[ht]
  \centering
  \includegraphics[width=\columnwidth]{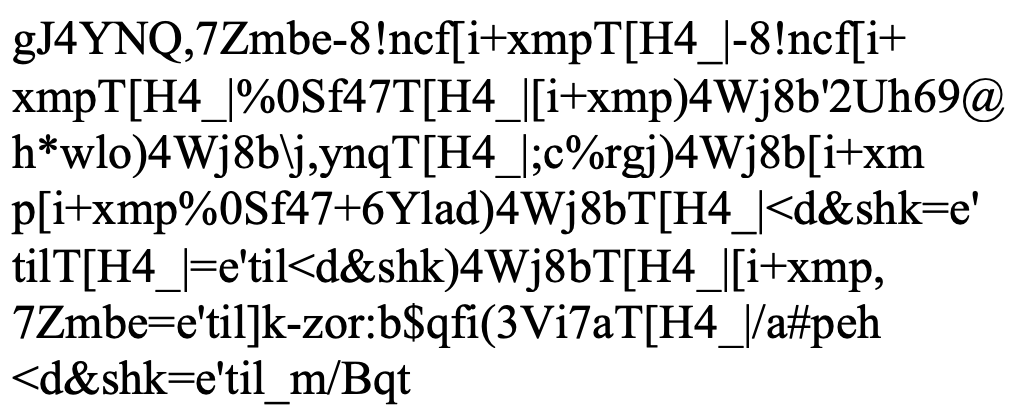}
  \caption{Ciphertext for Kasiski Analysis}
  \label{fig:ka}
\end{figure}
Initially, the occurrence of similar character sequences in our ciphertext is recorded. In this case, the character `4' appears 16 times in the ciphertext at positions 2, 27, 45, 52, 57, 67, 85, 99, 109, 130, 139, 147, 165, 181, 189, and 231. Subsequently, the distances between these occurrences and the first occurrence are calculated and result in 25, 43, 50, 55, 65, 83, 97, 107, 128, 137, 145, 163, 179, 187, and 229 respectively. After calculating the Greatest Common Divisor (GCD) of the distances, which is 1 in this case, the resulting value likely represents the key length used for generating the ciphertext. However, this value does not match the actual key length in this case.

Our experimentation with the Kasiski test unveiled intriguing outcomes. The test displayed noteworthy accuracy in predicting key lengths less than 3, achieving accurate estimations. However, as the key length exceeded this threshold, the precision of predictions notably declined, surpassing the traditional ciphers. Consequently, this impedes the efficient application of frequency analysis on the substrings produced through key-length division.

\subsection{
Entropy Analysis }
The calculation of Shannon entropy using (\ref{eq:entropy}) \cite{shannon} provides insights into the degree of disorder and unpredictability within encrypted messages. The entropy formula is expressed as follows:

\begin{equation}
\textit{Entropy} = -\sum_{i} P(x_i) \log_2 P(x_i)
\label{eq:entropy}
\end{equation}

Where $P(x_i)$ represents the probability of occurrence of character $x_i$. 
\begin{figure}[ht]
  \centering
  \includegraphics[width=\columnwidth]{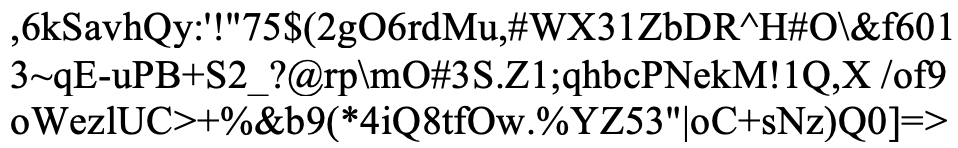}
  \caption{Ciphertext for Entropy Analysis}
  \label{fig:ea}
\end{figure}

In our experimentation, an entropy value of approximately 6.495 was obtained for the ciphertext provided in Fig \ref{fig:ea}, with a fixed key length of 6. This high entropy value indicates a substantial level of randomness and unpredictability within the ciphertext, suggesting a stronger level of security against attacks attempting to exploit patterns in encrypted data.

\subsection{Resilience to Large Language Models}
 A state-of-the-art Large Language Model (LLM) named Bidirectional Encoder Representations from Transformers (BERT) was employed\cite{LLm}. The model was subjected to training using a dataset comprising 1000 pairs of plaintext-ciphertext samples, each intricately linked to a corresponding key length. The primary objective was to determine whether the plaintext could be adeptly predicted by BERT through extrapolation from the ciphertext and key length, effectively reversing the intricate encryption process. While impressive accuracy was demonstrated by BERT for key length 1, a significant decrease in performance was observed with the increase in key length. This decline can be attributed to the inherently randomized nature of our cipher, which poses a challenge to BERT's cryptanalysis capabilities. Moreover, the augmented key length introduced heightened complexity, further testing the boundaries of BERT's analytical capabilities.

The robust and resilient nature of our proposed methodology in the modern world of cryptography is underscored by its resilience against contemporary cryptographic analysis techniques.

\section{RESULTS AND ANALYSIS}
\subsection{Noteworthy Observations}
Significant contributions to the advancement of traditional substitution ciphers are made. The key contributions can be summarized as follows:
\subsubsection{Unparalleled Keyspace Strength}

     With a staggering keyspace of \(4.40 \times 10^{25}\), the mere 25 possibilities of the Caesar cipher are vastly outperformed, and the impressive \(1.55 \times 10^{25}\) unique configurations that the Playfair cipher offers are even exceeded, establishing its advancement over traditional substitution techniques.
\subsubsection{Diverse Character Spectrum}
      A prominent role was assumed by special characters and numeric symbols, confounding conventional expectations of frequency analysis. Security is further bolstered by the nonlinear and nondeterministic nature of the proposed polyalphabetic encryption technique, as non-standard frequencies are introduced. Consequently, traditional frequency-based attacks, a common threat to substitution-based ciphers, are rendered significantly less effective.

  \subsubsection{Cryptanalysis Resilience}  
 Outperforming the traditional Vigenère cipher in the Kasiski test, and evaluated against an LLM like BERT, as key length increased, heightened complexity led to a noticeable decline in performance. This decrease in accuracy is attributed to the cipher's inherent randomized nature, which presents a challenge to BERT's cryptanalysis capabilities. The introduction of an augmented key length further escalated complexity.

\section{Conclusion}Classical substitution ciphers find revitalization through the integration of a randomized key matrix, effectively addressing the limitations of traditional methods while maintaining their simplicity and accessibility. Competitive computational speeds are not only achieved by the proposed methodology but also by the demonstration of remarkable resilience against frequency-based attacks, thus mitigating the vulnerabilities that substitution ciphers face. The extensive keyspace offered by the method serves as a robust defense mechanism against exhaustive search attacks like brute force. Additionally, the adaptation of the method to contemporary demands is showcased by its resilience to LLM, such as BERT.

Looking ahead, promising avenues for further exploration and application of the technique emerge. The optimization of key sizes to suit diverse application scenarios stands as a key direction for future research, enabling the tailoring of the method to specific security requirements. Further research may aim to ensure that the size of the ciphertext does not exceed that of the plaintext. Furthermore, the real-world implementation of the method in contexts such as low-powered sensors and offline encryption will bridge the gap between theoretical understanding and practical deployment, thus validating its efficacy in real-time scenarios.  
\bibliography{references.bib}

\end{document}